\documentstyle[12pt]{article}
\begin{document}
\def\mR{${\bf R^4}$}
\def\R{\bf R^4}
\def\mRn{${\bf R^n}$}
\def\Rn{\bf R^n}
\def\mRx{${\bf R^4_\Theta}$}
\def\Rx{\bf R^4_\Theta}
\def\msx{$\bf\Sigma^7_\Theta$}
\def\sx{\bf\Sigma^7_\Theta}
\def\xt{\times_\Theta}
\def\dM{${\cal D}$}
\def\dMx{${\cal D}_\Theta$}
\def\bdM{${\bar{\cal D}}$}
\def\bdMx{${\bar{\cal D}}_\Theta$}
\def\DS {differentiable structure}
\def\DSs {differentiable structures}
\def\SS {smoothness structure}
\def\SSs {smoothness structures}
\def\CS {complex structure}
\def\CSs {complex structures}
\newcommand{\statem}[2]{\begin{center}\vspace{\baselineskip}
\parbox{.85\textwidth}{{\bf #1:}\ {\em
#2}}\vspace{\baselineskip}\end{center}}

\begin{center}
{\bf EXOTIC SMOOTHNESS ON SPACETIME
}
\vskip 1cm
Carl H. Brans
\footnote{e-mail: brans@beta.loyno.edu} 
\vskip 1cm
{\it
Loyola University, \\
New Orleans, LA 70118, USA \\
}
\end{center}
\vskip 1cm
\noindent {\bf Abstract}
\vskip 0.7cm
Recent discoveries in differential topology are reviewed in light of 
their possible implications for spacetime models and related  subjects
in theoretical physics.  Although not often noted, a particular 
smoothness (differentiability) structure must be imposed on a
topological manifold before geometric or other structures of 
physical interest can be discussed.  The recent discoveries of 
interest here are of various surprising ``exotic'' smoothness 
structures on topologically trivial manifolds such as ${S^7}$ 
and ${\bf R^4}$. Since no two of these are diffeomorphic to each 
other, each such manifold represents a physically distinct model
of topologically trivial spacetime.  That is, these are not merely 
different coordinate representations of a given spacetime. The path  to  
such structures intertwines many branches of mathematics and theoretical physics 
(Yang-Mills and other gauge theories). An overview of these 
topics is provided,  followed by certain results concerning the 
geometry and physics of such manifolds.  Although exotic 
${\bf R^4}$'s cannot be effectively exhibited by finite 
constructions, certain existence and non-existence results can be 
stated. For example, it is shown that the ``exoticness'' can be 
confined to a time-like world tube, providing a possible model for an
exotic source. Other suggestions and conjectures for future research
are made. 
\vskip 1cm
\noindent {\bf I. Introduction}
\vskip 0.7cm

In the general context of
mathematical physics, ``exotic''  might refer to a class of
mathematical facts  that are 
surprising and highly counter-intuitive.  Such features
are often discovered in
the construction of counter examples to assumptions that seem
very reasonable, especially to physicists. In the specific sense 
intended in the title of this paper, the word applies to 
non-standard smoothness structures on topologically simple 
spaces, such as Milnor spheres and
 \mRx, denoting a smooth manifold 
 homeomorphic, but not 
diffeomorphic, to standard \mR.
   By way of 
preparation for the study of
these topics, we review some other surprising,
``exotic'' facts in the topology and \CS\ of 
otherwise simple spaces.
    \par
  Progress in theoretical physics has often come as a 
result of questioning old assumptions, e.g., \par
\begin{enumerate}
\item{\it spacetime should be an absolute product, 
time $\times$ space,}
\item{\it spacetime should be 
geometrically flat,}
\item{\it spacetime should have trivial topology,}   
\par
\end{enumerate}
and many others.  Questioning these natural assumptions 
obviously has led to many rich discoveries.   The Galilean       
structure of space and time in Newtonian physics was based on 1),
which certainly seems ``natural'' from everyday experience. Of 
course, we now know from special relativity that such a product 
structure is not absolute but relative to the state of motion of 
the observer.  Even granted such special relativistic insights, 
the geometric triviality of space, if not of spacetime, also
seems to be an inevitable consequence of experience.  The 
questioning of 2) however, led to the magnificent theory of 
general relativity. In hindsight, questioning of assumption 3) 
now seems to be part of a natural progression, and indeed, much 
work in modern theoretical physics calls  on non-trivial 
topological models. \par
In this questioning spirit then,  
 it would seem to be well worthwhile to  explore
the recent discovery of exotic \DSs\ on 
topologically trivial spaces, especially \mR.  Almost all 
widely investigated physical theories make use of differential 
equations which of necessity require a manifold with 
such a structure.  Of course, {\it locally}, all such structures 
are equivalent, so that the form of the equations and the local 
behavior of their solutions will be unchanged.  Nevertheless, 
{\it globally}, the \DSs\ are not equivalent, so neither is the 
underlying physics.  That is, such studies lead to fields that 
cannot be globally physically equivalent to any studied to date,
and may offer a rich resource of new physical possibilities.
\par
 We start with a brief discussion of what structures are really 
needed to do physics on a spacetime model, trying to make
explicit any hidden assumptions. Next,
 we consider some 
 more easily accessible examples of exotica: 
Weierstrass functions, \CSs\ on ${\bf R^2}$, 
Whitehead spaces, and Milnor spheres.
Then we
 explore some of the foundations of {\bf differential 
topology}, which, until recently,
 has generally been assumed to have only trivial
implications for physics, and follow this with
 an overview of
some of the highlights in the discovery of 
exotic \mR's. Finally we survey
some results of possible geometrical and physical significance
that have been obtained for \mRx's. \par
At this point it might be helpful to
review some obvious pro's and con's for the study of exotica
in theoretical physics.  The pro's are mostly mine and
undoubtedly reflect my own personal prejudice and experience.
These might include:
\par
\begin{itemize}
\item The
subject  involves beautiful and exciting math (it's fun!).
\item There is rich historical precedent, i.e., 
the investigation of what was at one time ``exotic'', such as
non-euclidean geometry and topology, has turned out to be 
fundamentally important for physics.
\item The product decomposition  time $\times$ space is not 
absolute in the physical sense and it turns out that
the non-triviality of
  this decomposition is very interesting even at 
the topological level (Whitehead spaces).
\item The discovery of exotic \DSs\ has involved a 
great deal of mathematics closely associated with physics,
such as
 Yang-Mills theory, gauge theory, peculiar dimension four, 
etc.
\item Exotic \mRx's present an infinity of previously 
unexplored, 
topologically trivial but physically inequivalent 
four-dimensional spacetime models.
\end{itemize}\par
On the con side, I am indebted to anonymous referees and other
skeptical colleagues for pointing out the following:
\begin{itemize}
\item The exotic \mRx's are nothing physically
 new, they merely provide``other'' manifolds whose physical 
 significance must be demonstrated.
\item Physicists
shouldn't waste time on any manifold unless an Einstein
metric can be displayed on it.
\item
Such studies are only abstract mathematics.
\end{itemize}\par
I am tempted to add another personal speculation 
on the objectors:
\begin{itemize}
\item (Inferred) ``I don't understand the subject.''
\end{itemize}\par
\vskip 1cm
\noindent{\bf II.  Spacetime Structures}
\vskip 0.7cm
To do physics, we need some model of space time.  
Clearly such a model must provide at least the following
features:\par
\begin{itemize}
\item {\bf Point set.} Thus, individual real and possible
``events'' must each have their own unique identity. Quantum
measurement theory clearly calls this assumption into question,
but no widely accepted, feasible, alternative seems to exist
at present.
\item
{\bf Topology.} The notion of convergence of events to a limit
seems to be a necessary precursor to the subsequent structures,
but again, quantum theory calls the observability of such
structures into serious question.
\item
{\bf  Smoothness, differentiability, $C^\infty$.}  
This is clearly needed for describing the differential equations 
that have been indispensable for physics since Newton invented
calculus.
\item {\bf Geometry, bundle structures, etc.}  These 
 provide the top level of structures needed for contemporary
physical theories.
\end{itemize}
Until now, the middle transition, 
$$point\ set\rightarrow topological\ space 
\buildrel ?? \over \rightarrow
smooth\ manifold \rightarrow bundles, etc.$$
was thought to be fairly well understood,  explored,
and trivial.
However, this assumption has recently been proven wrong by
 the discovery of 
 {\bf exotic smoothness} on topologically trivial \mR,
which will form the central part of our discussion
in the following.  \par
\vskip 1cm
\noindent{\bf III. Easily Accessible Exotica (Toy Models)}
\vskip 0.7cm
Even though the topology of \mR\ is about as simple as could be 
imagined, the problem of exotic \DSs\ on it involves some
very deep and difficult results in differential topology not 
readily accessible to most theoretical physicists.   So, by
way of analogy, we begin with a review of several ``toy'' models 
of unusual or exotic structures that can be understood 
in terms of more familiar mathematics.  \par
\vskip 1cm
{\bf 1. Weierstrass Functions as Exotica}
\vskip  0.7cm
A naive conjecture from elementary calculus is that every
function which is continuous over some interval must be at least 
piecewise smooth, i.e., its derivative exists except at isolated 
points.  ``Physical'' intuition might well suggest that this 
conjecture is valid.  However, it is not, as demonstrated by the
very nice ``Weierstrass'' functions, such as
\begin{equation}
W(t)=\sum_0^\infty a^k \cos(b^kt),\label{w1}
\end{equation}
where $\vert a\vert<1.$  Clearly, this series is absolutely 
convergent to a continuous function for all t.  However, naive 
term by term differentiation under the summation results in
\begin{equation}
W^\prime(t)\buildrel ??\over = -\sum_0^\infty 
(ab)^k\sin(b^kt).\label{w2}
\end{equation}
If $\vert ab\vert$ is chosen to be greater than one, the 
convergence of this series is dubious at best.  In fact, it can 
be shown rigorously\cite{strom}
 that the derivative of $W(t)$ does not 
exist 
anywhere.  This is a excellent counter example to the excessive
use of ``physical'' intuition in calculus.  In fact, graphing
various finite sum approximations to 
equations (\ref{w1}) and (\ref{w2}) provides even more insight.
\par
\vskip 1cm
{\bf 2. Complex Structures on ${\bf R^4}$ as Toy Physics}
\vskip 0.7cm
Complex structures are much more ``rigid'' than \DSs,
 and thus more easy to classify.
 Consider the case of establishing a \CS\ on ${\bf R^2}$.
The standard one is generated by one neighborhood with
\begin{equation}
(x,y)\rightarrow x+iy\in {\bf C^1}.\label{cs1}
\end{equation}
In this case, diffeomorphisms are replaced by {\bf 
biholomorphisms}.  Consider a different \CS, 
\begin{equation}
(x,y)\rightarrow x-iy.\label{cs2}
\end{equation}
This is certainly different, but the homeomorphism, 
$(x,y)\rightarrow (x,-y)$ is actually a biholomorphism, so these 
two are complex equivalent.  However, it is easy to construct 
another one which is not biholomorphic to the standard one, thus 
an {\bf exotic} \CS.  For example, let $(x,y)\rightarrow 
(g_x,g_y)$ be some homeomorphism of the plane into the open unit 
disk and define a second \CS\ by
\begin{equation}
(x,y)\rightarrow g_x+ig_y.\label{cs3}
\end{equation}
This second structure
cannot be biholomorphic (equivalent) to the standard \CS,        
equation (\ref{cs1}),
since there are no bounded non-constant holomorphic functions in 
the standard structures, but many in the new one.\par
Recall that the vacuum two-dimensional Maxwell electrostatic
equations are equivalent to the condition that $E_x-iE_y$
be a holomorphic function of the underlying complex variable.
If the complex structure is changed from the standard one,       
equation (\ref{cs1}), to the different one, equation (\ref{cs2}),
the underlying physics is not changed, since there is a 
biholomorphism of the plane on itself which makes the two 
descriptions of possible fields equivalent.  However, if the 
``exotic,'' non-standard \CS\ described by equation (\ref{cs3}) 
is 
chosen, then the physics is changed, since for this second one 
there will be non-constant, but bounded, electric fields.
  Thus, in some sense,  there would be different           
``physics'' resulting from the choice of the exotic \CS.  
Certainly this result is not to be taken  seriously as
physics, but
it does supply some motivation for suspecting that there may
be  material of potential physical significance  hiding
in mathematical structures which are
generally restricted only to  some ``standard'' types.
\vskip 1cm
{\bf 3. Some Exotic Topological Products}
\vskip 0.7cm
Another class of non-intuitive results in low dimensions is 
provided by {\bf Whitehead spaces}.  These models are
topological ones, and don't require the imposition
of any smoothness. 
However, the result of Moise mentioned below
shows that these spaces have unique 
smoothness anyway.
\par
 A Whitehead space, $W$, is an open, 
contractible three-dimensional topological manifold which has the 
following exotic properties:\par
\begin{equation}
W\ne {\bf R^3},\label{wh1}
\end{equation}
but
\begin{equation}
{\bf R^1}\times W={\bf R^4}.\label{wh2}
\end{equation}
In other words,
{\it
 it is not correct to assume that
when an ${\bf R^1}$
 is factored in \mR\ the result will necessarily be ${\bf   
R^3}$.}\par
This too is a profoundly counter-intuitive result. The 
construction of Whitehead spaces can be visualized 
using an infinite sequence of twisting tori inside each other.
The limit of the infinite iteration of this process produces a 
set whose complement in ${R^3}$ is a Whitehead space.  What the 
implications of this construction are for the smooth case are not 
now fully understood, but seem to be highly intriguing.
Newman and Clarke \cite{nc} have considered such structures 
in the context of the Cauchy problem in spacetime.  However, it 
should be pointed out that this paper refers to an alleged 
``proof'' of the Poincar\'e conjecture, which turns out not to 
have been valid after all.
\vskip 1cm
{\bf 4. Milnor Spheres}
\vskip 0.7cm
Fortunately there are a class of manageable exotic 
 structures available in the 
{\bf  smooth} category. These were
  discovered in the early 60's by Milnor \cite{mi}
  The simplest one is an exotic $S^7$.  This space can be 
realized naturally as the bundle space of an $SU(2)\approx S^3$ 
bundle over $S^4$ (which is compactified ${\bf R^4}$) using a 
construction of Hopf.    From the 
physics viewpoint, a Yang-Mills field with appropriate asymptotic 
behavior is a cross section of such a principal bundle.  
Such fields satisfying Yang-Mills field equations are 
called {\bf instantons} and turn out to be important later in the
story of exotica.   For now, however, consider the construction 
of $S^7$ as the subset of quaternion 2-space, $\{(q_1,q_2):\vert 
q_1\vert^2+\vert q_2\vert^2=1\}$.  There is a natural projection 
of this space into projective quaternion space, $(q_1:q_2)$.
This space, however, turns out to be nothing more than $S^4$.  
The kernel of this map is the set of unit quaternions, 
$S^3\approx SU(2)$.  Equivalently, $S^7$ can be defined by two
copies of $({\bf H}-0)\times S^3$, with identification
$$(q,u)\sim (q/\vert q\vert^2,qu/\vert q\vert)$$
Milnor was able to generalize this to produce a manifold, 
$\Sigma^7$ by means of the identification
$$(q,u)\sim (q/\vert q\vert^2,q^juq^k/\vert q\vert)$$
Milnor then  showed that if $j+k=1$ the space $\Sigma^7$ 
is {\it topologically} identical (homeomorphic) to $S^7$.  
However, if $(j-k)^2$ is not equal to 1 mod 7, then $\Sigma^7$ is 
{\bf exotic}, that is not diffeomorphic to standard $S^7$.  
\vskip 1cm
\noindent{\bf IV. Differential Topology = Global Calculus}
\vskip 0.7cm
In defining any point set, $X$,
 there may not be  a priori any
 way to associate numbers with a given point,
${\bf p}\in X$.  For spacetime models, {\bf p} is a
physical, real or possible, event.   The basic tools for 
analyzing space and time are associated with the notions of
Cartesian geometry.  In this approach, the set of events
is assumed to be numerically describable.  From the physical
viewpoint,
the process of assigning numbers is 
associated with a {\bf reference frame}, mathematically
by a {\bf coordinate patch}.  \par
It is easy to get lazy and falsely secure about this matter since
most spaces, $X$, considered in both physics and mathematics are
modelled by subsets of \mRn, so each {\bf p} is ``naturally''
associated, at least locally, 
 with  ordered sets of real numbers. 
However, it is
well known  that
{\it the definition of coordinates is not unique.}
This fact is of course 
at the very heart of the 
principle of general relativity:
\statem{Question} {Does re-coordination have 
any physical consequences?}
The investigation of the mathematical and physical consequences 
of assigning coordinates to  abstract points or physical events
logically begins at the {\bf topological} level, making the set,
$X$ into a {\bf topological manifold}.  Such a set is a 
topological space, covered by a family of open sets, 
an {\bf atlas of charts}, together with homeomorphisms,
maps $(U,\phi_U:U\rightarrow \Rn)$.    In other words, a 
topological
 manifold is one which is locally equivalent to \mRn\ in the
 topological sense.\par
However, in order to
do calculus, we need to use these local coordinates, say 
$x^\alpha$, to define derivatives, $\partial f/\partial 
x^\alpha.$  If more than one coordinate patch is needed, then
we must require
 differential consistency in overlaps, 
 \begin{equation}
\phi_U\cdot\phi^{-1}_V\in C^\infty,\label{ds1}
\end{equation}
where the combined map is from one open set in \mRn\ to another
so that the usual notion of smoothness or differentiability,
$C^\infty$, is well defined.  Such a smoothly consistent
family, ${\cal U}=\{U,\phi_U\},$ is called a {\bf smooth atlas}.
Another atlas, ${\cal U'}$ is consistent with ${\cal U}$ if and 
only if their union is again a smooth atlas.  Clearly the set of
such atlases is a partially ordered set and any one then defines
a maximal one.   
  A {\bf \DS}\ 
on $X$ is defined by such
a maximal atlas, and makes $X$ into a {\bf differentiable}, or 
{\bf smooth manifold}.  Clearly any atlas consistent with
the maximal one defines it, so \DSs\ are usually defined
by less than the maximal one.  For example, for \mRn, 
the atlas can be defined by only one set, $U=\Rn,\phi_U={\bf 1}$.
The resulting \DS\ on \mRn\ is called the standard one.
From the physical viewpoint a \DS\ is necessary to do
 calculus over $X$, and thus is
 obviously indispensable for the definition of any physical
theory. The mathematical discipline dedicated to the study
of smooth manifolds is called {\bf differential topology}.
Two excellent texts on the subject are by Hirsch \cite{h}
and by Br\"ocker and J\"anich \cite{bj}.
\par
 A \DS\  contains coordinate transitions not only
{\it within} a given  $X$,
 but also allows the definition of a
natural {\it equivalence} established by a {\bf diffeomorphism}.
This is a homeomorphism of one smooth manifold to another (or 
itself), $f:X\rightarrow Y,$ which together with its inverse is
smooth when expressed in the atlases on $X$ and $Y$ respectively.
As general relativity has grown in allowing arbitrary smooth     
manifolds to serve as spacetime models, the notion of physical
reference frame transformations has been associated with the
global recoordination that is a diffeomorphism.  Thus,
\statem{Principle of General Relativity}{The laws
of physics should be formally invariant under recoordination,
and the diffeomorphism group defines this
natural equivalence class for physics.}
Similarly, the diffeomorphism group forms the natural equivalence
class for the mathematics of differential topology. 
In the following we will at times loosely misuse the term ``\DS''
on a given $X$ for a diffeomorphism equivalence class of such.
  From the
mathematical viewpoint then, a
fundamental problem is whether or not such an equivalence class 
is trivial for a given {\it topological} $X$.  That is
\statem{Question}{Can a given topological space support
 truly distinct, non-diffeomorphic, \DSs?  Or,
can two non-diffeomorphic smooth manifolds be homeomorphic? }
The issue raised 
in this question is a subtle one and very easy to misunderstand 
so let us begin by examining
 the difference between {\bf 
different}, and {\bf non-diffeomorphic} structures on a given
topological manifold, using  the real line as an example.\par
Thus, take
$X ={\bf R^1}=\{\bf p\}$, each 
element being a single real number. From this comes
the ``natural'' smoothness structure, 
${\cal D}_1$, generated from one coordinate
patch, $U=X,$ and
\begin{equation}
\phi_{1}({\bf p})=p,\label{r11}
\end{equation}
  that is, the         
coordinate is simply the numerical value associated with the
topological point, {\bf p}.
Similarly, consider two others, ${\cal D}_2,{\cal D}_3$,
generated also from one patch, with the same domain, $U=X,$
but with 
\begin{equation} 
 \phi_{2}({\bf p})=2p,\label{r12}
 \end{equation}
and
\begin{equation}
\phi_{3}({\bf p})=p^{1/3}.\label{r13}
\end{equation}
  Clearly, ${\cal D}_1$ is not {\bf different} from ${\cal D}_2$
since the maximal atlases generated by both are the same, in 
fact,
\begin{equation}
\phi_{2}\cdot\phi^{-1}_{_1}(p)=2p\in C^\infty.\label{r14}
\end{equation}
Nevertheless they are both {\bf different} from ${\cal D}_3,$ 
since the coordinates are incompatible in the overlap:
\begin{equation}
\phi_{3}\cdot\phi^{-1}_{_1}(p)=p^{1/3}\notin 
C^\infty.\label{r15}
\end{equation}
The important point however, is that these {\bf different}
structures, ${\cal D}_1$ and ${\cal D}_3$,
 are in fact {\bf diffeomorphic}, and thus equivalent
from the viewpoint both of physics and of
 the mathematics of differential 
topology.  The diffeomorphism is established with the {\bf 
homeomorphism,} $f:{\bf p}\rightarrow{\bf p}^3,$ so
\begin{equation}
\phi_{3}\cdot f\cdot
\phi^{-1}_{_1}(p)=\phi_{3}(f({\bf p}))=(p^3)^{1/3}=
p\in C^\infty.\label{r16}
\end{equation}
  In fact, 
\statem{Fact}
{Any two \DSs\ on ${\bf R^1}$ are 
diffeomorphic to each other.}
In other words, there is essentially only {\it one} \DS\ that can 
be put on ${\bf R^1}$ both mathematically and physically.
The uniqueness of the smoothness structure on ${\bf R^1}$ 
 is probably not too surprising.  In fact
it can be generalized to\par
\statem
{Theorem(Moise)}{There is one and only one \DS 
\ on any topological manifold of dimension $n<4$.}\par
The case of higher dimensions cannot be settled so generally.
However, using Thom cobordism techniques, the special cases
of the topologically trivial ${\bf R^n}, n>4$ can be,\par
\statem
{Theorem} {There is one and only one \DS\ on \mRn\ for $n>4$,
namely the standard one.}\par
The standard cobordism results are not applicable for
the $n=4$ case, so it remained as an open question
until the early 80's and the arrival of \mRx's.
\statem
{Open Question as of the early 80's} 
 {Are there \DSs\ on \mR which are not diffeomorphic to 
   the standard one?}\par
\vskip 1cm
\noindent{\bf V. The Road to Exotica on \mR.}
\vskip 0.7cm
The work of many people, using tools from
various fields of mathematics  and theoretical physics, 
began to suggest an affirmative answer to this question,
 and
 it is now known that there are in fact an uncountable
infinity of inequivalent (non-diffeomorphic) \DSs\ on topological 
\mR.  
The discovery of exotic 
smoothness 
on topological \mR's, producing manifolds denoted by \mRx, 
involved developments from many branches of mathematics, 
including topology and differential equations.
Important results in this search were
  based on the study of {\bf moduli spaces}
derived from the physical model of Yang-Mills 
fields, that is non-Abelian gauge theory.  First recall that
a moduli space is built from 
a space of fields, ${\cal A}$,
 often gauge potentials, over a particular
manifold, $M$. Typically, these fields are further required to
satisfy certain field equations and to behave a certain way under
gauge transformations,
 ${\cal G}$.  In general ${\cal A}$ will be a huge set,
certainly not a finite dimensional manifold.  So, how can
moduli spaces be managed?
It turns out that when the
gauge transformations are factored out, the result
\begin{equation}
{\cal M}={\cal A}/{\cal G},\label{ms1}
\end{equation}
can be a well behaved space such as a finite dimensional 
manifold, perhaps with singularities.  ${\cal M}$ is a {\bf 
moduli space}.  As a simple example, consider the family of 
p-forms over a compact manifold, $M$.  Let the field equations be 
the restriction that the forms be closed.  Let the action of the
gauge group be the addition of an exact form.  The resulting 
${\cal M}$ in this case is just the $p^{th}$ deRham cohomology 
group, which is typically a finite dimensional vector space.
This is only a simple, not realistic example.  More productive is 
the study of instantons over $S^4$, which
are certain cross sections  
of the Hopf bundle, $S^7$, as investigated by 
{\bf Atiyah} and others.  For an excellent review of the 
techniques involved in studying moduli spaces, see 
Freed and Uhlenbeck \cite{fu}.\par
These studies lead to:
\statem{Fact} 
{The moduli space of certain fields
over a manifold can give information
about the differential topology of the manifold.}
This fact turned out to be of key importance in the road to the 
discovery of \mRx.  {\bf Donaldson} used moduli space studies to 
show that spaces with certain intersection forms (a topological
feature) could not be smoothed. For a four manifold, there is a 
natural symmetric bilinear form induced by the Poincar\'e
pairing of closed two-cycles. If the coefficient group is 
${\bf 
R}$, this can be understood in a dual sense as
\begin{equation}
(\alpha,\beta)=\int \alpha\wedge\beta,\label{dr1}
\end{equation}
for two closed two-forms.  However, for our applications the 
coefficient group must be the integers mod 2, so the form must be 
understood in terms of the geometric intersection of the two 
two-simplices contributing to the homology classes.  For example, 
for $S^2\times S^2$, the intersection form, is
\begin{equation}
\omega=\pmatrix{0 & 1\cr 1 & 0\cr}.\label{if1}
\end{equation}
For our purposes, Donaldson's contribution can be summarized by
a remarkable theorem coupling topology to smoothness.  By 
studying the moduli space of Yang-Mills fields over a smooth four 
manifold, $M$, Donaldson showed
\statem{Theorem (Donaldson)}
{If $M$ is smooth, simply connected, closed, and has definite
intersection form, $\omega$, then $\omega=\pm {\bf 1}.$}
Now, let $E_8$ stand for a certain standard $8\times 8$ 
symmetric integral form (actually associated with the exceptional
Lie algebra denoted by the same symbol), which happens not 
to be diagonal.  Then Donaldson's theorem clearly results in
\statem{Corollary}
{If the intersection form of $M$ is $-2E_8$, $M$ cannot be given
a smooth structure.}
\par
  {\bf Freedman} 
built on Casson
handlebody construction and other techniques to derive an
important result which we can summarize as\par
\statem{Theorem (Freedman)}{If a topological four manifold, $X$,
 is non-compact, simply connected, has $H_2(X,{\bf 
Z})=0$, and a single
  end  collared 
by ${\bf R}\times S^3$, then $X$ is homeomorphic to \mR.}\par
Note that this is a purely {\it topological} result, but coupled 
with Donaldson's results will lead to \mRx.\par
Next, consider the {\bf Kummer surface,} $K$, which is the subset 
of $CP^3$ defined by 
\begin{equation}
K=\{(z_1:z_2:z_3:z_4)\vert z_1^4+z_2^4+z_3^4+z_4^4=0\}.\label{K1}
\end{equation}
It turns out that the intersection form for $K$ is
\begin{equation}
\omega_K=-2E_8+3H,\label{K2}
\end{equation}
where $H$ is the intersection form for $S^2\times S^2$ described 
in (\ref{if1}), and $3H$ is that of the connected sum of three 
such 
spaces.  Using {\it topological} (not smooth) surgery, the $3H$ 
part of $\omega_K$ can be ``localized''
as  $3\#S^2\times S^2,$ collared to the remainder, $K'$ by a 
topological ${\bf R}\times S^3.$ The critical question now is
whether or not this is a {\it smooth} product.  If it were, it 
could be smoothly capped by a four-disk to produce
  a smooth
four-manifold with $\omega=-2E_8,$ violating Donaldson's 
corollary stated above.\par
Consequently, the collar must be exotic ${\bf R}\xt S^3,$
{\it not} diffeomorphic to the standard smooth product.  Smoothly
capping one end of this results in a manifold satisfying the
topological conditions of Freedman's theorem above, and thus is
homeomorphic to \mR.  However, the exotic nature of the end 
implies that it cannot be {\it diffeomorphic} to the standard
smooth version of this space, and so must be exotic \mRx.\par
After this initial discovery, rapid progress was made in 
constructing (more in the sense of existence) various \mRx's, and
classifying them. For example, Gompf has a paper entitled
``An Exotic Menagerie,''\cite{g2}, showing the existence of an 
uncountable number of non-diffeomorphic \mRx's.  
Gompf's construction makes extensive use of handlebody 
chains, which apparently must be infinite.
Freedman and
Taylor \cite{ft} show the existence of a universal \mRx\ in which
all others can be smoothly embedded.  Also, as a note for use 
below, it turns out that some \mRx's can be smoothly embedded in
standard \mR, and others cannot.\par
Recently,  field equations suggested by {\bf Seiberg}
 and {\bf Witten} \cite{sw}
  show great promise for simplifications
of the study of moduli spaces.  
 However, to date,  it is unfortunately true that
\statem{Fact}
{No finite effective coordinate patch presentation
exists of any exotic \mRx.}
Nevertheless, 
even in the absence of a manageable coordinate patch 
presentation, certain features can be explored.  Some are 
summarized in results from previous papers.\par
\vskip 1cm
\noindent{\bf VI. Some Geometry and Physics on \mRx.}
\vskip 0.7cm
Even though an explicit, effective coordinate patch 
presentation of \mRx\ is not available, certain additional facts 
about such space, including some of a
 geometric and thus physical sort can be discovered.  For a more 
complete 
exposition and discussion of these results see \cite{cb1}
and \cite{cb2}.  Here we merely review some of the results.\par
First, the question naturally arises concerning the given global
topological coordinates, $\{p^\alpha\}$, which define the 
topological manifold \mR, and their relationship to
 the local smooth coordinates
given by the coordinate patch functions, $\phi_U^\alpha$.
Both provide maps from an abstract ${\bf p}\in \R,$ 
into \mR\ itself.  Clearly the global topological coordinates
cannot themselves be smooth everywhere since otherwise they
would provide a diffeomorphism of \mRx\ onto standard \mR.
But can they be locally smooth?  This is answered in the 
affirmative by
\statem{Theorem}
{There exists a smooth copy of each \mRx\ for which the global 
$C^0$ coordinates are smooth in some neighborhood.  That is, 
there exists a smooth copy, 
${\bf R^4_\Theta}=\{(p^\alpha)\},$ for which
$p^\alpha\in C^\infty$ for $\vert{\bf p}\vert<\epsilon.$}
The implied obstruction to continuing the $\{p^\alpha\}$ as 
smooth beyond the $\epsilon$ limit presents a challenging source 
for further investigation.  Related to this is a the defining 
feature of the early discovery work of \mRx's, namely the
non-existence of arbitrarily large smoothly embedded 
three-spheres.\par
There are also certain natural ``topological but not smooth''
decompositions.  For example,
\statem{Theorem}
{\mRx\ is the topological, but not smooth,
 product, ${\bf R^1}\xt{\bf R^3}$.}\par
Many interesting examples can be constructed using Gompf's
``end-sum'' techniques \cite{g1}. In this construction           
topological ``ends'' of non-compact smooth manifolds are 
glued together smoothly, $X\cup_{end}Y.$
  If one of the manifolds, say $X$, is also
topological \mR, the topology of the resultant space is 
unchanged, that is ${\bf R^4}\cup_{end}Y$ is homeomorphic to $Y$.
  However, if $X$ is an \mRx\ which cannot be smoothly embedded 
in standard \mR, then neither can the the end sum.  Thus,
\statem{Gompf's end sum result}
{If $X=\Rx$ cannot be smoothly embedded in standard \mR,
but $Y$ can be, then
${\bf R^4_\Theta}\cup_{end}Y$ is homeomorphic, but not
 diffeomorphic to $Y$.  }
 This technique will be used further below.\par
To do geometry we need a metric of the appropriate signature.  
It is a well known fact that any smooth manifold can be endowed 
with a smooth Riemannian metric,
$g_0$.  This follows from basic bundle 
theory \cite{st}.  Similarly, if the Euler number of $X$ vanishes
a globally non-zero smooth tangent vector, $u$ exists.  
$g_0$ and $u$ can be combined then to construct a global smooth
metric of Lorentz signature, $(-,+,+,+)$, in dimension four.
A generalization of this result follows also from standard bundle
theory, \cite{st}.
\statem{Theorem}
{If $M$ is any smooth connected 4-manifold and $A$ is a closed
submanifold for which
$H^4(M,A; {\bf Z})=0,$ then any smooth
time-orientable Lorentz signature metric defined
over $A$
can be smoothly continued to all of $M.$}
One immediate conclusion about certain geometries on \mRx\ 
 can be drawn from an investigation of the exponential map
of the tangent space at some point, which is standard \mR,
onto the range of the resulting geodesics. 
  The Hadamard-Cartan theorem guarantees
that this map will be a diffeomorphism onto the full manifold
if it is simply connected, the geometry has nonpositive curvature
and is geodesically complete\cite{ghl}. Thus,
\statem{Theorem}
{ There can be no  geodesically complete Riemannian
metric  with nonpositive sectional curvature on
\mRx.}
The apparent lack of localization of the ``exoticness'' means
that it must extend to infinity in some sense as illustrated by 
the lack of arbitrarily large smooth three-spheres.  However, it 
turns out to be possible that the exoticness can be localized in 
a {\it spatial} sense as follows:
\statem{Theorem}
{There exists smooth manifolds which are homeomorphic but not
diffeomorphic to ${\bf R^4}$ and for which the global topological
coordinates $(t,x,y,z)$ are smooth for
$x^2+y^2+z^2\ge \epsilon^2>0,$ but not
globally.  Smooth metrics exists for which the boundary of this region is
timelike, so that the exoticness is spatially confined.}
The details of the construction of such manifolds are given in 
\cite{cb1}.  First, Gompf's end-sum technique is used to produce 
a \mRx\ for which the global topological coordinates are smooth 
outside of the cylinder, that is, in the closed set
 $c_0=\{(t,x,y,z)\vert 
x^2+y^2+z^2\ge\epsilon\}$
 described in the first part of the 
theorem.  Next, a  Lorentz signature metric
 is constructed on $c_0$.  This metric can even
be a vacuum Einstein metric.  The only condition is that the 
$\partial/\partial t$ be time like on $c_0$.  The cross section
continuation result with $A=c_0$ then guarantees the extension
of the metric over the full space consistent with the conditions 
of the theorem. What makes the complement of $c_0$ exotic is the 
fact that the $(x,y,z,t)$ cannot be continued as smooth functions 
over all of it.  This result leads to
\statem{Conjecture}
{This localized exoticness can act as a source for some externally regular
field, just as matter or a wormhole can.}\par
Another set of interesting physical possibilities arise in a 
cosmological context inspired by the exotic product,$X={\bf 
R}\xt S^3$, which arises from a puncturing of \mRx.  It is not
hard to apply the same techniques used above to show that this
product can be the standard smooth one for a finite, or 
semi-infinite range of the first variable, say $t$.  The 
resulting manifold could then be endowed with a standard 
cosmological metric.  This metric, and even the variable $t$ 
itself, cannot be continued as globally smooth indefinitely, 
because of the exotic smoothness obstruction.  Recall, however, 
that $X$ is still a globally smooth manifold, with some globally 
smooth Lorentz-signature metric on it.  Other interesting 
topological but not smooth products can be constructed by use of 
the end-sum construction.  One interesting example is exotic 
Kruskal, $X_K={\bf R^2\times_\Theta S^2}$.  Using the cross
section continuation theorem above, the standard vacuum Kruskal 
metric can be imposed on some closed set, $A\subset X_K$, and 
then continued to some smooth metric over the entire space.  
However, it cannot be continued as Kruskal, since otherwise
$X_K$ would then be standard ${\bf R^2}\times S^2$.
In sum,\
\statem{Theorem}
{ On some smooth manifolds
which are topologically
${\bf R^2\times S^2}$, the standard Kruskal metric
cannot be smoothly continued over the full range, $u^2-
v^2<1.$}\par
Finally, we close this section with a brief mention of the 
possible physical significance of Milnor's exotic seven-spheres, 
\msx.  Recall that the standard $S^7$ as a Hopf bundle is the 
underlying bundle space for Yang-Mills ($SU(2)$) connections 
over $S^4$, which is compactified \mR.  On the other hand, 
\msx\ as a bundle is no longer a principle $SU(2)$ bundle, but 
one associated to a principle $Spin(4)$ bundle.  This could be 
regarded as some sort of generalized or exotic Yang-Mills bundle. 
It might prove interesting to investigate the possible physical 
ramifications of this.
\vskip 1cm
\noindent{\bf VII. Conclusion and Conjectures}
\vskip 0.7cm
From the principle of general relativity as generally defined, we 
learn that two different smooth manifolds can represent the same 
physics, merely presented in different coordinate representation, 
if and only if they are diffeomorphic to each other.  Until 
recently, this diffeomorphism class has been regarded by 
physicists as relatively trivial and the construction of ``new'' 
spacetime models seemed to require changes of the basic 
topology.  From this review, however, it is apparent that this is 
not the case, that there are an infinity of physically 
inequivalent representations of spacetime all having the trivial 
topology of the first model, \mR.  It would   seem very 
surprising, and contrary to much historical precedent, to have 
the  sudden and unexpected discovery of the
richness of mathematical models for 
four dimensional spacetime to be of no physical significance at 
all.  

\end{document}